\documentstyle[epsf,preprint,aps,eqsecnum]{revtex}
\newcommand{\Ymark}
{{\normalsize\boldmath $Y\hspace{-1mm}ukawa~
I\hspace{-.3mm}n\hspace{-.3mm}stitute~K\hspace{-.5mm}yoto$}
\hfill {\rm\normalsize YITP-96-21}\\~ \\}
\newcommand{\ket}[1]{|#1\rangle}
\newcommand{\ketm}[1]{\ket{\mu_{(#1)}}}
\newcommand{\sgn}[1]{{\rm sgn}(#1)}
\newcommand{\rap}[2]{k_{#1_{#2}}^{(#2)}}
\newcommand{\qnum}[1]{I_{a_{#1}}^{(#1)}}
\newcommand{\ch}{\!\!\circ\!\!}
\begin{document}
\draft
\preprint{}
\title{\Ymark SU($\nu$) Generalization of Twisted 
Haldane-Shastry Model}
\author{ Takahiro Fukui\cite{Ema,Padd}} 
\address{Yukawa Institute for Theoretical Physics, 
Kyoto University, Kyoto 606-01, Japan}
\author{Norio Kawakami}
\address{Department of Applied Physics,
and Department of Material and Life Science,\\
Osaka University, Suita, Osaka 565, Japan}
\date{July 3, 1996}
\maketitle
\begin{abstract}
The SU($\nu$) generalized Haldane-Shastry spin chain
with $1/r^2$ interaction is studied with twisted boundary conditions.
The exact wavefunctions of Jastrow type 
are obtained for every rational value of the twist angle 
in unit of $2\pi$. The spectral flow of the 
ground state is then discussed as a function of the twist angle.
By resorting to the motif picture in the 
Bethe ansatz method, we show that the period of the spectral
flow is $\nu$, which is determined by 
the statistical interaction in exclusion statistics.
\end{abstract}
\pacs{PACS numbers: 75.10.Jm, 05.30.-d\\
Keywords: SU($\nu$) Haldane-Shastry model, Twisted boundary
condition, Spectral flow, Asymptotic Bethe ansatz, Fractional
exclusion statistics, Yangian symmetry} 
\section{introduction}
The Haldane-Shastry spin chain 
\cite{Hal,Sha} is a one-dimensional integrable lattice 
model with $1/r^2$ interaction \cite{ISMD1,ISMD2}, 
which has provided a number of valuable notions
in low-dimensional quantum systems.
Among others, (fractional) exclusion statistics
proposed by Haldane \cite{HalFES} has received 
considerable attention. From  the viewpoint of exclusion statistics,
the Haldane-Shastry model is regarded as an idealized model 
which is completely free from any irrelevant perturbations and
describes the fixed point Hamiltonian for 
massless spin chain models. In this idealized circumstance, 
it may be  possible to observe characteristic
 features inherent in exclusion statistics.

Motivated by this, we have recently extended 
the Haldane-Shastry model to include twisted 
boundary conditions, and have shown that exclusion 
statistics can be explicitly observed in the 
period of the spectral flow\cite{FukKaw2}.
The strategy to probe exclusion statistics has been based on 
the idea that 
the response to twisted boundary conditions 
(or equivalently to external gauge fields) enables
us to extract the knowledge 
on various properties for interacting particle systems
\cite{ABB,ShaSut,Sut,FukKaw1,KusAok}.
As mentioned above, the Haldane-Shastry model 
is free from any irrelevant perturbations, allowing us to 
observe exclusion statistics clearly in the spectral flow.
It has been indeed shown\cite{FukKaw2} 
that the period of the ground state is solely
determined by the statistical interaction 
in exclusion statistics.

In this paper, we generalize the twisted Haldane-Shastry model 
to the case with SU($\nu$) spin symmetry, and solve it exactly.
Such a generalization to the multicomponent models
may be interesting in its own right
\cite{HaHal,Kaw} and also its close relationship to
the fractional quantum Hall effect.
We show that the period of the spectral flow in the SU($\nu$)
model is indeed determined by the matrix of the statistical 
interaction in exclusion statistics,
i.e. the SU($\nu$) Cartan matrix.

In the next section, we introduce the Hamiltonian of the SU($\nu$)
Haldane-Shastry model with twisted boundary conditions.
We explain how the twisted boundary condition is imposed
consistently with the long-range nature of the $1/r^2$ interaction.
In section III, the exact solution of the 
twisted SU($\nu$) model is derived by exploiting the Jastrow-type
ansatz. In section IV, the exact spectrum thus obtained is shown 
to be reproduced by the Bethe ansatz.
We discuss the spectral flow of the energy spectrum
as a function of the twist angle in section V. 
By resorting to the notion of the motif \cite{HHTBP} 
in the Bethe ansatz, we give
an interpretation for the period of the spectral flow 
in terms of exclusion statistics.
Summary and discussions are given in section VI.

\section{model hamiltonian}

The Haldane-Shastry model has long-range $1/r^2$ 
interaction, so that it is not trivial to 
impose twisted boundary conditions without
loss of integrability. 
 Recently, we have proposed one of the ideas for this purpose
\cite{FukKaw2}. We first recall that  
the Haldane-Shastry model\cite{Hal,Sha} was originally introduced 
 by imposing the periodic boundary condition and by summing up
the pair-wise $1/r^2$ interaction around the ring 
infinite times \cite{ISMD2}.  From this point of view,
we start with  the following Hamiltonian,
\begin{eqnarray}
H&=&\sum_{n<n'}\sum_{m=-\infty}^{\infty}\frac{1}{(n-n'-mN)^2}
\frac{1}{2}P_{n,n+mN}\nonumber\\
&=&\sum_{n<n'}\sum_{m=-\infty}^{\infty}\frac{1}{(n-n'-mN)^2}\nonumber\\
&&\qquad\times\left\{
\frac{1}{2}\sum_{\alpha\in\Phi_+}
\left(
E_n^\alpha E_{n'+mN}^{-\alpha}+E_n^{-\alpha}E_{n'+mN}^\alpha
\right)
+\left(\sum_{k=1}^{\nu-1}
H_n^kH_{n'+mN}^k+\frac{1}{2\nu}
\right)
\right\},
\label{Ham1}
\end{eqnarray}
where $n$ denotes the index for sites $n=1,2,\cdots,N$ and
$P_{n,n'}\equiv 2\sum_aT_n^aT_{n'}^a+1/\nu$ is the exchange operator in
SU($\nu$) algebra.
In the second line,
$\Phi_+$ is the set of the positive roots of SU($\nu$) algebra, 
the generator $E_n^\alpha$ is the so-called step operator 
associated with the root $\alpha$ at site $n$, and
$H_n^k$ with $k=1,2,\cdots,\nu-1$
is the hermitian generator in Cartan subalgebra.
We fix the normalization of the generators as
${\rm tr}(T^aT^b)=\frac{1}{2}\delta_{ab}$.
The Hamiltonian (\ref{Ham1}) was already solved in the case of
the periodic boundary condition
\cite{HaHal,Kaw}, where the summation over $m$ can be
explicitly carried out to give the well-known 
sine-inverse-square interaction \cite{ISMD2}.
However, once we impose twisted boundary conditions,
it becomes a non-trivial solvable model compatible with 
twisted boundary conditions. 
The form of the Hamiltonian (\ref{Ham1}) implies that the
Cartan basis may be a natural basis to treat the system 
with twisted boundary conditions.
By this reason, we shall span the basis of the wave function
in terms of the Cartan basis, which is different 
from that in \cite{HaHal}, though both bases give the same 
results in the case of the periodic boundary condition.

To be more specific in our notations, let us denote the 
simple roots by $\alpha_{(i)}$ with $i=1,2,\cdots,\nu-1$.
Then the positive roots can be expressed as
$\alpha_{(k)}+\cdots+\alpha_{(l)}$
with $1\le k\le l\le\nu-1$. Step operators associated with the 
positive roots are denoted simply as 
$E^k=E^{\alpha_{(1)}+\cdots+\alpha_{(k)}}$ for $k=1,\cdots,\nu-1$ and
$E_n^{(k,l)}\equiv E_n^{\alpha_{(k+1)}+\cdots+\alpha_{(l)}}$
for $1\le k<l\le\nu-1$.
We label the state at each site as  $\ket{\mu_{(i)}}$,
which is specified by the $\nu$ weight vectors 
$\mu_{(i)}$ with $i=0,1,\cdots,\nu-1$
in the fundamental representation.
As we have fixed the normalization of the generators, the weight
vectors satisfy 
\begin{equation}
\mu_{(i)}\cdot\mu_{(j)}=\left\{
\begin{array}{ll}\frac{1}{2}-\frac{1}{2\nu}&{\rm for~}i=j\\
-\frac{1}{2\nu}&{\rm for~}i\ne j.\end{array}\right.
\end{equation}
The highest weight state is assumed to 
be $\ket{\mu_{(0)}}$, and other
descendant states are created by the lowering operators $E^{-k}$ such
that $\ket{\mu_{(k)}}=E^{-k}\ket{\mu_{(0)}}$.
Therefore, in this representation,
$E^{k}$  connects directly the highest 
weight state $\ketm{0}$ with a descendant 
state $\ketm{k}$, while $E^{(k,l)}$ connects two 
descendant states $\ketm{k}$ and $\ketm{l}$.

The boundary condition we impose on the Hamiltonian (\ref{Ham1}) 
in this paper is
\begin{eqnarray}
&&E^{\pm k}_{n+mN}=e^{\pm 2\pi i\phi_km}E^{\pm k}_n,\nonumber\\
&&E^{\pm(k,l)}_{n+mN}=e^{\pm 2\pi i(\phi_l-\phi_k)m}E^{\pm(k,l)}_n,\nonumber\\
&&H^k_{n+mN}=H^k_n,
\label{BouCon}
\end{eqnarray}
where the first line is the definition of the $\nu-1$ twist angles 
$\phi_k$ and the second line naturally follows from the commutation
relation of the step operators.
In what follows, we refer to the angle $\phi_k$ defined in unit of
$2\pi$ as the twist angle.
Note that we can impose twisted boundary 
conditions independently on each species.


In order to obtain the exact solution, 
it is convenient to introduce the 
periodic operators by the following 
gauge transformations 
\begin{eqnarray}
&&E_n^{\pm k}\rightarrow e^{\pm2\pi i\phi_kn/N}E_n^{\pm k},\nonumber\\
&&E_n^{\pm(k,l)}\rightarrow 
e^{\pm2\pi i(\phi_k-\phi_l)n/N}E_n^{\pm(k,l)}.
\label{trans}
\end{eqnarray}
In the remainder of the paper, we always 
use the step operators in the gauge transformed form.
Now we restrict ourselves to  rational values
\begin{equation}
\phi_k=\frac{p_k}{q_k}
\end{equation}
with $p_k$ and $q_k$ being integers.
This constraint is, at present, essential to solve the twisted model.
Then the summation over $n$ can be done explicitly, 
and the Hamiltonian (\ref{Ham1}) is now written as
\begin{equation}
H=\left(\frac{\pi}{N}\right)^2\frac{1}{2}(T+H_{\rm int}),
\label{FinHam}
\end{equation}
where $T$ and $H_{\rm int}$ are the hopping and interaction terms,
respectively. The former is defined by
\begin{equation}
T=\sum_{k}\sum_{n\ne n'}
J_{\phi_k}(n-n')E_n^kE_{n'}^{-k}+
\sum_{k<l}
\sum_{n\ne n'}
J_{\phi_l-\phi_k}(n-n')E_n^{(kl)}E_{n'}^{-(kl)}.
\end{equation}
Here $J_\phi(n)$ is the $1/r^2$ coupling compatible with 
the boundary condition (\ref{BouCon}),
\begin{equation}
J_\phi (n)\equiv\frac{1}{q^2}\sum_{m=0}^{q-1}
e^{2\pi ip(n+mN)/qN}\sin^{-2}\left[\frac{\pi (n+mN)}{qN}\right],
\quad{\rm for}~\phi=\frac{p}{q}.
\end{equation}
For later convenience, we rewrite the above expression as follows:
\begin{equation}
T=\sum_{k}T_k(\phi_k)
+\frac{1}{2}
\sum_{k\ne l}T_{kl}
(\phi_{kl}),
\label{TotHop}
\end{equation}
where $\phi_{kl}=\phi_k-\phi_l$ and
\begin{equation}
T_k(\phi)=\sum_{n\ne n'}
J_\phi(n-n')E_n^kE_{n'}^{-k},\quad
T_{kl}(\phi)=U_lT_kU_l^\dagger .
\label{FinHop}
\end{equation}
Here $U_l$ is  the unitary operator of the $\pi$ rotation around
$\alpha_{(1)}+\cdots+\alpha_{(k)}$, i.e.,
$U_k=\exp\{i\frac{\pi}{2}(E^k+E^{-k})\}$.
The interaction term is
\begin{equation}
H_{\rm int}=\sum_{n\ne n'}J_0(n-n')
\left(\sum_{k=1}^{\nu-1}
H_n^kH_{n'}^k+\frac{1}{2\nu}
\right).
\end{equation}
Note that $J_0(n)$ is of the usual sine-inverse-square form
$J_0(n)=\sin^{-2}\frac{\pi n}{N}$.
This completes the introduction of the 
 twisted SU($\nu$) Haldane-Shastry model.
We can see that after the gauge transformation
(\ref{trans}) the effect of twisted boundary conditions is 
incorporated in effective hopping, 
and hence  we can solve the Hamiltonian with 
periodic boundary conditions.


\section{exact solution}

In this section we  construct the exact eigenstates 
for the model (\ref{FinHam}) and obtain the 
corresponding eigenenergies. In the previous paper\cite{FukKaw2}, 
we have solved the simplest case, i.e. the twisted SU(2)
model, by using the Jastrow-ansatz wave function.
In that case, we have demonstrated  
that the Jastrow wave function can still be an eigenfunction of
the twisted Haldane-Shastry model if we convert the 
operators into the periodic ones via the gauge transformation.
In what follows, we will show that this is also the case for
the SU($\nu$) model, and then obtain its exact solution.

Before starting the calculation, let us fix some notations.
There are $\nu$ different states at each site.
We take the highest weight state $\ket{\mu_{(0)}}$ 
 as  the reference state (background), and regard other 
descendant states $\ketm{k}$ ($k=1,2,\cdots \nu-1$)
as particle states.
Since the number of states in $\ketm{k}$ is conserved,
let us denote it as $M_k$.
Then the relation $\sum_{k=0}^{\nu-1}M_k=N$ holds.
Let the subset of the sites denoted by 
$\{n^{(k)}_{\alpha_k}\}$ with $\alpha_k=1,\cdots,M_k$
be the positions of the state $\ketm{k}$.

We now propose the following Jastrow-ansatz state as a 
candidate for the exact eigenstate 
of the twisted SU($\nu$) model,
\begin{equation}
\ket{\Psi}=\sum_{k=1}^{\nu-1}\sum_{n_1^{(k)}<\cdots<n_{M_k}^{(k)}}
\psi(\cdots,\{n^{(k)}\},\cdots;\cdots,J_k,\cdots)
\prod_{k=1}^{\nu-1}\prod_{\alpha_k=1}^{M_k}E_{n_{\alpha_k}^{(k)}}^{-k}
\ket{\mu_0,\mu_0,\cdots,\mu_0},
\end{equation}
where $\psi$ is a Jastrow-type wave function
\begin{equation}
\psi=\prod_{k=1}^{\nu-1}\prod_{\alpha_k=1}^{M_k}
z^{J_kn_{\alpha_k}^{(k)}}
\prod_{k=1}^{\nu-1}\prod_{\alpha_k<\beta_k}
d(n_{\alpha_k}^{(k)}-n_{\beta_k}^{(k)})^2
\prod_{k<l}\prod_{\alpha_k,\alpha_l}
d(n_{\alpha_k}^{(k)}-n_{\alpha_l}^{(l)}).
\label{Jas}
\end{equation}
Here, $d(n)=\sin(\pi n/N)$, $z=\exp(2\pi i/N)$, and 
the current for each species is defined by 
\begin{equation}
J_k=\frac{N-M_k-M_0}{2} \quad {\rm mod~}1.
\label{Cur}
\end{equation}
We wish to prove below 
that this is indeed an eigenstate of the Hamiltonian.


Let us start by evaluating the action of the hopping terms on the
wavefunction. We have introduced two kinds of the hopping Hamiltonians 
$T_k(\phi_k)$ and $T_{kl}(\phi_{kl})$ in eq.(\ref{FinHop}).
The former exchanges the pairs of $\{n^{(0)}\}$ and $\{n^{(k)}\}$, 
while the latter exchanges the pairs of $\{n^{(k)}\}$ 
and $\{n^{(l)}\}$.
In what follows, we first evaluate the action of $T_k$ on the wave
function, which can be calculated directly.
Although the direct manipulation of $T_{kl}$ turns out to be difficult,
there is a remarkable trick to simplify the calculation,
found by Wang, Liu and Coleman\cite{WLC}
for the supersymmetric $t$-$J$ model.
By generalizing this trick to the SU($\nu$) model, we 
then evaluate the action of $T_{kl}$ by the use of 
the results obtained for the action of $T_k$.

Although the wavefunction (\ref{Jas}) contains the
coordinates of various species, we note that
$T_k$ acts only on $\ket{\mu_{(0)}}$ and $\ket{\mu_{(k)}}$.
We thus classify the set of total sites into 
three subsets $\{n^{(0)}\}$, $\{n^{(k)}\}$ and the remainder 
$\{n\}-\{n^{(0)}\}-\{n^{(k)}\}$.
To simplify the expressions, 
we introduce the following notations
for the elements of each subset,
\begin{eqnarray}
&&x_\alpha\in\{n^{(k)}\},\quad \alpha=1,2,\cdots,M_k\nonumber\\
&&y_j\in\{n\}-\{n^{(k)}\}-\{n^{(0)}\},\quad j=1,2,\cdots,N-M_k-M_0.
\label{SimNot}
\end{eqnarray}
Then, the problem reduces essentially to 
that of two species with the background, and one 
finds that the action of $T_k$ on (\ref{Jas}) is 
analogous  to that 
of the spin-hopping term for the supersymmetric 
$t$-$J$ model with $1/r^2$ interaction\cite{FukKawTJ}. 
Therefore, by exploiting the techniques developed there,
we have
\begin{equation}
\frac{T_k(\phi_k)\psi}{\psi}=
\sum_{n=1}^{N-1}J_{\phi_k}(n)z^{J_kn}
\sum_{\alpha=1}^{M_k}\prod_{\beta(\ne\alpha)=1}^{M_k}
B_{\alpha\beta}^{(n)}
\prod_{j=1}^{N-M_k-M_0}F_{\alpha j}^{(n)},
\label{TBF}
\end{equation}
where
\begin{equation}
B_{\alpha\beta}^{(n)}=1-g_{\alpha\beta}^{(n)}, \quad 
g_{\alpha\beta}^{(n)}=
\frac{(1-z^n)z^2_\alpha+(1-z^{-n})z^2_\beta}{(z_\alpha-z_\beta)^2}
\end{equation}
with $z_\alpha=z^{x_\alpha}$ and
\begin{equation}
F_{\alpha j}^{(n)}=\cos\frac{\pi n}{N}
+\sin\frac{\pi n}{N}\cot\Theta_{\alpha j},\quad
\Theta_{\alpha j}=\frac{\pi(x_\alpha-y_j)}{N}.
\label{Thelm}
\end{equation}
The last factor $F$ in eq.(\ref{TBF}) describes the interaction
between different species.
Note that such interaction terms automatically disappear
in the case of SU(2). 
Expanding the above formula in power series of $z^n$ and $z^{-n}$, 
and using the original notation for sites,
we finally find the expression for the action of $T_k$,
\begin{equation}
\frac{T_k(\phi_k)\psi}{\psi}=\sum_{i=1}^4W_i^{(k)},
\label{ActTk}
\end{equation}
where
\begin{eqnarray}
&&W_1^{(k)}=2M_k\widetilde\varepsilon(J_k+\phi_k)
+\frac{1}{3}M_k(N^2-1)+\frac{1}{6}M_k(M_k^2-1)+
\frac{1}{2}M_kM_0^2,\\
&&W_2^{(k)}=\frac{1}{2}\sum_{\alpha_k\ne\beta_k}
J_0(n_{\alpha_k}^{(k)}-n_{\beta_k}^{(k)})
-\frac{1}{2}\sum_{\alpha_k,\alpha_0}
J_0(n_{\alpha_k}^{(k)}-n_{\alpha_0}^{(0)})
+\frac{1}{2}\sum_{\stackrel{\scriptstyle{m=1}}{(m\ne k)}}^{\nu-1}
\sum_{\alpha_m,\alpha_k}
J_0(n_{\alpha_k}^{(k)}-n_{\alpha_m}^{(m)}),
\label{TwoBod1}\\
&&W_3^{(k)}=2i\left(J_k+\phi_k-\frac{N}{2}\right)
\sum_{\alpha_k,\alpha_0}\cot\Theta_{\alpha_k\alpha_0},
\label{TwoCur1}\\
&&W_4^{(k)}=\sum_{\alpha_k\ne\beta_k,\alpha_0}
\cot\Theta_{\alpha_k\beta_k}\cot\Theta_{\alpha_k\alpha_0}
+\sum_{\alpha_0\ne\beta_0,\alpha_k}
\cot\Theta_{\alpha_0\beta_0}\cot\Theta_{\alpha_0\alpha_k}.
\label{ThrBod1}
\end{eqnarray}
The function $\widetilde\varepsilon (J+\phi)$ is 
\begin{equation}
\widetilde\varepsilon(J+\phi)
\quad=\left\{
\begin{array}{ll}\varepsilon(J+\phi),&~\mbox{\rm for~ integer}~J,\\
\frac{1}{2}\left(\varepsilon(J+\phi-\frac{1}{2})
+\varepsilon(J+\phi+\frac{1}{2})-\frac{1}{2}\right),&
~\mbox{\rm for~  half-integer}~J,
\end{array}\right.
\end{equation}
where $\varepsilon (k)$ is defined for every rational $k$ as 
\begin{equation}
\varepsilon(k)=
\left(2[k]-N+1\right)k-[k]\left([k]+1\right).
\label{SinParEne}
\end{equation}
The expression (\ref{ActTk})
holds for the currents which satisfy
\begin{equation}
\frac{N+M_k-M_0}{2}-1\le J_k+\phi_k
\le N-\left(\frac{N+M_k-M_0}{2}-1\right).
\end{equation}
The derivation of the above equations is outlined in Appendix.

We now move to the action of $T_{kl}$.
As already mentioned, 
$T_{kl}$ exchanges the pairs of $\{n^{(k)}\}$ and $\{n^{(l)}\}$,
making its action on the wave function
(\ref{Jas}) rather complicated. 
However, we note the following identity
\begin{eqnarray}
&&\psi(\cdots,\{n^{(k)}\},\cdots;J_1\cdots,J_k,\cdots J_{\nu-1})\nonumber\\
&&=A\psi(\cdots,\{n^{(0)}\},\cdots;
J_1-J_k+N/2,\cdots,N-J_k,
\cdots,J_{\nu-1}-J_k+N/2),
\end{eqnarray}
where $A$ is a constant independent of coordinates.
Let us denote the r.h.s. of the above equation as $A\psi_{(k)}$. 
Combining these two relations, we can evaluate the action of $T_{kl}$
quite similarly to that of $T_k$.  The key formula is
\begin{equation}
\frac{T_{kl}(\phi_{kl})\psi}{\psi}=
\frac{T_{k}(\phi_{kl})\psi_{(l)}}{\psi_{(l)}},
\end{equation}
which was previously found for the  supersymmetric $t$-$J$ model
\cite{WLC}.
The relation implies that what we have to do is to 
substitute the following replacements to eq.(\ref{ActTk}),
\begin{equation}
\{n^{(0)}\}\leftrightarrow\{n^{(l)}\}\quad(M_0\leftrightarrow
M_l),\quad  
J_k\rightarrow\left\{
\begin{array}{cc}J_{kl}+\frac{N}{2}&{\rm for}~k\ne l\\
N-J_k&{\rm for}~k=l,\end{array}\right.
\end{equation}
where $J_{kl}=J_k-J_l$. The results are
\begin{equation}
\frac{T_{kl}(\phi_{kl})\psi}{\psi}=\sum_{i=1}^4W_i^{(kl)},
\label{ActTkl}
\end{equation}
where
\begin{eqnarray}
&&W_1^{(kl)}=2M_k\widetilde\varepsilon(J_{kl}+N/2+\phi_{kl})
+\frac{1}{3}M_k(N^2-1)+\frac{1}{6}M_k(M_k^2-1)+
\frac{1}{2}M_kM_l^2,\\
&&W_2^{(kl)}=\frac{1}{2}\sum_{\alpha_k\ne\beta_k}
J_0(n_{\alpha_k}^{(k)}-n_{\beta_k}^{(k)})
-\frac{1}{2}\sum_{\alpha_k,\alpha_l}
J_0(n_{\alpha_k}^{(k)}-n_{\alpha_l}^{(l)})
+\frac{1}{2}\sum_{\stackrel{\scriptstyle{m=0}}{(m\ne k,l)}}^{\nu-1}
\sum_{\alpha_m,\alpha_k}
J_0(n_{\alpha_k}^{(k)}-n_{\alpha_m}^{(m)}),
\label{TwoBod2}\\
&&W_3^{(kl)}=2i\left(J_{kl}+\phi_{kl}\right)
\sum_{\alpha_k,\alpha_l}\cot\Theta_{\alpha_k\alpha_l},
\label{TwoCur2}\\
&&W_4^{(kl)}=\sum_{\alpha_k\ne\beta_k,\alpha_l}
\cot\Theta_{\alpha_k\beta_k}\cot\Theta_{\alpha_k\alpha_l}
+\sum_{\alpha_l\ne\beta_l,\alpha_k}
\cot\Theta_{\alpha_l\beta_l}\cot\Theta_{\alpha_l\alpha_k},
\label{ThrBod2}
\end{eqnarray}
provided that the currents satisfy the condition
\begin{equation}
\frac{M_k-M_l}{2}-1\le J_{kl}+\phi_{kl}\le 1-\frac{M_k-M_l}{2}.
\end{equation}
Consequently,
 from eqs.(\ref{ActTk}) and (\ref{ActTkl}), we have the action of
the total hopping Hamiltonian (\ref{TotHop}) on the wave function,
\begin{equation}
\frac{T\psi}{\psi}=\sum_{k=1}^4W_i,
\label{ActT}
\end{equation}
where 
\begin{equation}
W_i=\sum_{k=1}^{\nu-1}W_i^{(k)}+\frac{1}{2}
\sum_{\stackrel{\scriptstyle{k,l=1}}{k\ne l}}^{\nu-1}W_i^{(kl)},
\quad {\rm for}~i=1,2,3,4.
\label{W}
\end{equation}
In this expression, there still 
remain unwanted two- and three-body terms.
In what follows, we show how these unwanted terms indeed vanish.

First, we can easily confirm that 
the two-body terms depending on the currents, $W_3$, vanish.
The other two-body terms $W_2$ are reduced to
\begin{eqnarray}
W_2&=&\frac{1}{2}\sum_{k=1}^{\nu-1}\sum_{\alpha_k\ne\beta_k}
J_0(n_{\alpha_k}^{(k)}-n_{\beta_k}^{(k)})
-\frac{1}{2}\sum_{k=1}^{\nu-1}\sum_{\alpha_k,\alpha_0}
J_0(n_{\alpha_k}^{(k)}-n_{\alpha_0}^{(0)})\nonumber\\
&&\qquad\qquad\qquad\qquad+\frac{1}{4}(\nu-2)\sum_{k=1}^{\nu-1}
\sum_{\alpha_k}\sum_{n(\ne n_{\alpha_k}^{(k)})}
J_0(n_{\alpha_k}^{(k)}-n)\nonumber\\
&=&\frac{1}{2}\sum_{k=0}^{\nu-1}\sum_{\alpha_k\ne\beta_k}
J_0(n_{\alpha_k}^{(k)}-n_{\beta_k}^{(k)})
-\frac{1}{6}M_0(N^2-1)+\frac{1}{12}(\nu-2)\sum_{k=1}^{\nu-1}
M_k(N^2-1).
\end{eqnarray}
Next, the three-body terms $W_4$ are 
calculated as
\begin{eqnarray}
W_4&=&\sum_{\stackrel{\scriptstyle{k,l=0}}{k\ne l}}^{\nu-1}
\cot\Theta_{\alpha_k\beta_k}\cot\Theta_{\alpha_k\alpha_l}\nonumber\\
&=&\sum_{k=0}^{\nu-1}\cot\Theta_{\alpha_k\beta_k}
\left(\sum_{n(\ne n_{\alpha_k}^{(k)})}\cot\Theta_{\alpha_kn}
-\sum_{\gamma_k(\ne\alpha_k)}\cot\Theta_{\alpha_k\gamma_k}
\right)\nonumber\\
&=&-\sum_{k=0}^{\nu-1}\sum_{\alpha_k\ne\beta_k}
J_0(n_{\alpha_k}^{(k)}-n_{\beta_k}^{(k)})
+\frac{1}{3}\sum_{k=0}^{\nu-1}M_k(M_k^2-1).
\end{eqnarray}
Combining these formulae, we  end up with
\begin{equation}
\frac{T\psi}{\psi}=2E-\frac{1}{2}\sum_{k=0}^{\nu-1}
J_0(n_{\alpha_k}^{(k)}-n_{\beta_k}^{(k)}).
\end{equation}
Now, it is seen that 
the two-body terms are exactly canceled out with $H_{\rm int}$.
Consequently, we can  prove that 
$\psi$ is an eigenfunction of the Hamiltonian, and the
corresponding eigenenergy,
defined by $H\psi=\left(\frac{\pi}{N}\right)^2E\psi$, is
given by
\begin{eqnarray}
E&=&\sum_{k=1}^{\nu-1}M_k\widetilde\varepsilon(J_k+\phi_k)+
\frac{1}{2}\sum_{\stackrel{\scriptstyle{k,l=1}}{(k\ne l)}}^{\nu-1}
M_k\widetilde\varepsilon(J_{kl}+N/2+\phi_{kl})\nonumber\\
&+&\frac{1}{24}(\nu+4)\sum_{k=1}^{\nu-1}M_k(M_k^2-1)
+\frac{1}{8}\sum_{k=1}^{\nu-1}(N-M_k-M_0)M_k^2\nonumber\\
&+&\frac{1}{24}(3\nu-2)(N-M_0)(N^2-1)
+\frac{1}{4}(N-M_0)M_0^2+\frac{1}{6}M_0(M_0^2-1)-\frac{1}{12}M_0(N^2-1),
\label{ExaEne}
\end{eqnarray}
where the energy depends explicitly on $\{\phi\}$ 
via the first line of the expression. 
This is the exact eigenenergy for 
the SU($\nu$) twisted Haldane-Shastry model, which is the 
main result in this section.

It is now instructive to write down the ground state energy.
If we restrict ourselves to 
$N=\nu M$ with an even integer $M$ and
$0\le\phi_1\le\phi_2\le\cdots\le\phi_{\nu-1}\le 1$,
the ground state is realized by the choice
$M_0=M_1=\cdots=M_{\nu-1}$ and $J_1=J_2=\cdots=J_{\nu-1}$.
Then the ground state energy as a function of the twisted angle 
reads
\begin{equation}
E_{\rm g.s.}=M\sum_{k=1}^{\nu-1}(2k-\nu+1)\phi_k
-\frac{1}{12}\nu^2(\nu-2)M^3-\frac{1}{12}\nu(2\nu-1)M.
\end{equation}
Note that this is the ground state energy
for restricted twist angles, from which it is difficult to get 
full information on the the spectral flow beyond the restriction,
though it is possible in principle. 
The main difficulty
comes from the expression of the spectrum (\ref{ExaEne}) 
which contains the Gauss symbol (see the definition
of $\varepsilon(k)$ in (\ref{SinParEne})). 
Namely, whenever one of $\{\phi_k\}$ jumps over an integer
or other $\phi_j$'s, the energy has a singular 
cusp structure (see also Fig.1).
It is hence  not easy to pass through these cusps  and
trace the spectral flow. 
In order to overcome the difficulty and to discuss
the spectral flow correctly in the full region of the twist angles, 
it is necessary to resort 
to an alternative method, i.e. the 
 Bethe ansatz, which will be described  in the 
following section.

\section{Bethe ansatz approach}

Before discussing the spectral flow, 
we first show that the exact spectrum obtained 
in the previous section can be
 reproduced by the Bethe ansatz
(sometimes called the asymptotic Bethe ansatz\cite{ISMD2}).
The advantage to exploit the Bethe-ansatz description 
is, as will be shown below, that the spectral flow is
naturally determined by appropriately choosing the 
quantum numbers in the Bethe equations.
Also, by using the motif picture\cite{HHTBP}, which can 
visualize how the quantum numbers in the Bethe ansatz
are determined, we can naturally interpret the period of the
spectral flow in terms of fractional exclusion statistics.

In order to formulate the Bethe equations for the SU($\nu$) model,
let us introduce the $\nu-1$ kinds of rapidities, denoted here as
$k_{a_i}^{(i)}$ with $i=1,2,\cdots,\nu-1$, and
$a_i=1,2,\cdots,\sum_{\alpha=i}^{\nu-1}M_\alpha$.
Following a standard procedure in the nested
Bethe ansatz\cite{sunsuth}, we can formally write down
the Bethe equations as
\begin{eqnarray}
&&\rap{a}{1}=\qnum{1}+\phi_1-\frac{1}{2}\sum_{a_2}\sgn{\rap{a}{1}-\rap{a}{2}}
+\frac{1}{2}\sum_{b_1(\ne a_1)}\sgn{\rap{a}{1}-\rap{b}{1}}\nonumber\\
&&\hspace{5cm}\vdots\nonumber\\
&&\frac{1}{2}\sum_{b_i(\ne a_i)}\sgn{\rap{a}{i}-\rap{b}{i}}+\qnum{i}
+\phi_{i}-\phi_{i-1}=\nonumber\\
&&\hspace{3cm}\frac{1}{2}\sum_{a_{i-1}}\sgn{\rap{a}{i}-\rap{a}{i-1}}
+\frac{1}{2}\sum_{a_{i+1}}\sgn{\rap{a}{i}-\rap{a}{i+1}}\nonumber\\
&&\hspace{5cm}\vdots\nonumber\\
&&\frac{1}{2}\sum_{b_{\nu-1}(\ne a_{\nu-1})}
\sgn{\rap{a}{\nu-1}-\rap{b}{\nu-1}}+\qnum{\nu-1}
+\phi_{\nu-1}-\phi_{\nu-2}=
\frac{1}{2}\sum_{a_{\nu-1}}\sgn{\rap{a}{\nu-1}-\rap{a}{\nu-2}}
\label{BetAnsEqu}
\end{eqnarray}
with $i=2,\cdots,\nu-2$.
Here we note that the two-body phase shift
takes  the step-like form, which is inherent 
in the $1/r^2$ systems\cite{ISMD2}. This special form 
of the phase shift is essential for  the model to be  
ideal in the sense of exclusion statistics\cite{HalFES}.
The total energy of the model, defined in the unit of 
$\left(\frac{\pi}{N}\right)^2$ as before, is given by
\begin{equation}
E=\frac{1}{12}N(N^2-1)+\sum_{a_1=1}^{M_1+\cdots+M_{\nu-1}}
\varepsilon(\rap{a}{1})
\end{equation}
with $\varepsilon(k)$ defined in (\ref{SinParEne}).
The energy thus seems to be solely 
determined  by the rapidity $k^{(1)}$, but
we should remember that its configuration is affected by the other
rapidities via the nested equations.
Therefore, when we follow the spectrum as a function 
of the twist angle, we must specify
 the precise configuration including
$k^{(2)},\cdots,k^{(\nu-1)}$, whose
behavior should be quite different from each other.

To see clearly  how the rapidities are nested with
each other, we take the simplest but
non-trivial example, i.e. the SU(3) case.
It is  straightforward, though a little bit complicated, 
to generalize  the following discussions to the SU($\nu$) 
models. To be specific, we use the simplified notations, 
$k_{a_1}^{(1)}=k_\alpha$ and $k_{a_2}^{(2)}=\lambda_i$,
and write down the Bethe equations as
\begin{eqnarray}
&&k_\alpha=I_\alpha^{(1)}+\phi_1-\frac{1}{2}\sum_{\beta\ne\alpha}
\sgn{k_\alpha-k_\beta}
+\frac{1}{2}\sum_{j}\sgn{k_\alpha-\lambda_j},
\label{SU31}\\
&&\frac{1}{2}\sum_{j(\ne i)}\sgn{\lambda_i-\lambda_j}+I_i^{(2)}+\phi_{21}=
\frac{1}{2}\sum_\beta\sgn{\lambda_i-k_\beta},
\label{SU32}
\end{eqnarray}
where $\phi_{21}=\phi_2-\phi_1$.

Here we notice that eq.(\ref{SU32}),
which has been formally deduced by the nested Bethe ansatz,
 does not seem to hold for a fractional value 
of $\phi_{21}$ at a first glance.
This problem stems from the fact
that the two-body phase shift in the present system
is of the step-like form.
We now wish to explain how we can deal with eq.(\ref{SU32})
correctly.  To this end, let us 
consider the following configuration at 
$\phi_1=\phi_2=0$ as an example,
\begin{equation}
\cdots<k_{\alpha-1}<\lambda_{i}<k_{\alpha}<\cdots.
\end{equation}
If we put $\phi_{21}=1$, we see  from 
eq.(\ref{SU32}) that 
$I^{(2)}_i$ is shifted  
to $I^{(2)}_i+1$ and the above configuration
should be changed to
\begin{equation}
\cdots<k_{\alpha-1}<k_{\alpha}<\lambda_{i}<\cdots.
\end{equation}
Namely, the position of $\lambda_i$ is 
exchanged with that of $k_\alpha$ 
sitting on its right neighbor. 
The fractional $\phi_{21}$ between 0 and 1 
should smoothly interpolate these two configurations.
One readily notices that this is done by introducing an infinitesimal
width $\eta$ in the step-like phase shift and
then taking the limit of $\eta \rightarrow 0$. 
This observation naturally leads us to 
divide the l.h.s. of eq.(\ref{SU32}) as 
$I^{(2)}_i+\phi_{21}\equiv(I^{(2)}_i+[\phi_{21}])
+(\phi_{21}-[\phi_{21}])$. Then the fractional part 
$\phi_{21}-[\phi_{21}]$ can be  absorbed into  
$\sgn{\lambda_i-k_\alpha}\equiv\sgn{0}$ in the case of 
$\lambda_i=k_\alpha$ for $0<\phi_{21}<1$.
Consequently, we get the simple 
result from eq.(\ref{SU32})  
\begin{equation}
\frac{1}{2}\sgn{\lambda_i-k_\alpha}=\phi_{21}-\frac{1}{2}
\end{equation}
for $0<\phi_{21}<1$. By substituting the above formula 
into (\ref{SU31}), we finally have 
\begin{equation}
k_\alpha=\left\{
\begin{array}{ll}
k_{0\alpha}+\phi_1, \quad&
{\rm for}\quad\cdots<k_{\alpha-1}<k_\alpha<\cdots ,\\
k_{0\alpha}+\phi_2,\quad&
{\rm for}\quad\cdots<\lambda_i<k_\alpha<\cdots ,
\end{array}\right.
\label{ExpK}
\end{equation}
for $0\le\phi_{21}<1$, where $k_{0\alpha}$ is the rapidity
for $\{\phi\}=0$.  From this equation 
we can see that $k$ with (without) $\lambda$ 
in its left neighbor labels the rapidities for 
species-1 (-2) degree of freedom.  
Up to now, we have restricted ourselves to the case $\phi_{21}\le 0$.
For the other case $\phi_{21}\le 0$, $\lambda$ exchanges the position
with $k$ sitting on its left when $\phi_{21}$ decreases.
Namely, the role of ``left'' and ``right'' in the previous discussion
exchanges. 

The above interpretation of the two-body phase shift 
can be  generalized to the SU($\nu$) case. As a result, 
it is easily confirmed that  by suitably taking 
the quantum numbers $I^{(j)}_\alpha$, the Bethe equations 
(\ref{BetAnsEqu}) indeed reproduce the exact energy (\ref{ExaEne}) 
for the SU($\nu$) case obtained in the previous section.

\section{Spectral flow and exclusion statistics}

We now discuss the spectral flow with the aid of the Bethe ansatz
description.  For this purpose,
the notion of the motif is particularly useful, allowing
us to trace the spectral flow correctly.
Let us first recall that the solution of the Bethe equations
(\ref{BetAnsEqu}) can be  specified 
graphically by the sequence of 0 and 1
denoting respectively empty and occupied 
states of the momentum $k_j$\cite{HalFES}.
Note that the unoccupied momentum, 0, is introduced to 
represent  the repulsion effect of the two-body phase shift.
This is the essence of the motif.
For example, as was shown in ref. \cite{HalFES},
 the motif for the SU(2) case reads
$0101\cdots010$, which implies that the two-body phase shift
enlarges the spacing of rapidities twice as 
large as the free fermion case.
For the SU($N$) case, one can find
from the Bethe equations (\ref{BetAnsEqu}) 
that the rapidity $k_j^{(1)}$ is subject to 
the constraint that any configurations 
with more than $\nu-1$ consecutive 1's are prohibited.
For example, the ground state of the SU(3) chain described by 
eqs.(\ref{SU31}) and (\ref{SU32}) is
characterized by the motif for $k_\alpha$ 
\begin{equation}
011011\cdots0110 ,
\end{equation}
where 1 denotes the occupied $k_\alpha$.
In order to include the solution of the
auxiliary rapidity $\lambda_i$, the motif should read
more precisely
\begin{equation}
01\ch101\ch1\cdots01\ch10 ,
\end{equation}
where we have introduced   $ \, \ch \,$ 
which denotes the position of the occupied $\lambda_i$.
We need this type of the motif when we discuss the 
spectral flow for the SU($\nu$) case ($\nu \ge 3$)
with several different twist angles $\phi_i$.

To see how well the motif picture works for our problem, 
let us discuss the spectral flow 
by taking  the SU(3) model as an example.
In Fig.1, we have shown the numerical diagonalization 
results of the spectrum for the 
finite system.  These numerical results may be 
complementary to the exact results obtained in section III,
since the latter provides a specific series of the 
eigenstates. We can see several characteristic properties in the 
spectral flow, e.g. the linear $\phi$-dependence on the 
twist angel,  cusp structures, etc.  
These behaviors seem rather peculiar,
because the smooth $\phi^2$ dependence for small $\phi$
is expected for ordinary spin models with 
short-range interaction. We find that the above 
characteristic behaviors are closely related to high 
symmetry of the system which is essential for
the $1/r^2$ model to be an idealized model without
any irrelevant perturbations. We shall discuss later
that the linear $\phi$-dependence indeed results from high
symmetry of the $1/r^2$ model.

Now, let us analyze the spectral flow by exploiting the 
motif picture.  As mentioned above the motif for the 
SU(3) singlet ground state is given by
$0110110\cdots0110$. When the boundary is twisted as
$\phi_1=\phi_2=\phi$, 
the motif develops with the increase of the twist angle as 
\begin{equation}
0110110\cdots0110\rightarrow
1011011\cdots1010\rightarrow
1101101\cdots1100\rightarrow
0110110\cdots0110,
\label{motifsu3}
\end{equation}
where we have omitted $\, \ch \, $ for simplicity,
 because 1's move in parallel, and $\, \ch \, $
is always sandwiched by the same 1's.
Here, the arrow means that unit $\phi$ is added;
$\delta\phi=1$.  The corresponding spectral flow in Fig.1 
is: (a) $\rightarrow$ (b) $\rightarrow$ (c).
From this interpretation with the motif, 
we can say that the period of the above spectral flow is 3.
There may be another type of the  spectral flow,
since there are two independent twist angles $\phi_1$
and $\phi_2$.  For example, when $\phi_1=0$ and $\phi_2=\phi$,
the motif changes like,
\begin{eqnarray}
&&01\ch101\ch10\cdots01\ch10 \rightarrow 0101\ch101\cdots1\ch101\:\ch\:=
1\ch101\ch101\cdots1\ch100 \nonumber\\
\rightarrow&&101\ch101\ch1\cdots101\ch\:0=\:\ch\:101\ch101\cdots\:\ch1010
\rightarrow01\ch101\ch10\cdots01\ch10 ,
\end{eqnarray}
as the twist angle $\phi$  is increased.  The corresponding 
spectral flow thus has the period 3.
In both cases, we can see that the period of 
the spectral flow in the SU(3) model is 3.
Similarly we can analyze  the spectral flow  
for more general SU($N$) cases based on the 
motif. For example,  the period of the ground state 
turns out to be  $\nu$ for  
$\phi_1=\phi_2=\cdots=\phi_{\nu-1}=\phi$.
In this way, we can determine the period of the 
spectral flow in terms of the motif picture.

We now wish to discuss how the above 
description is related to the notion of
exclusion statistics.
According to Haldane\cite{HalFES}, the statistical 
interaction $g$  in exclusion statistics is defined as
\begin{equation}
\frac{\partial D_\mu}{\partial N_\nu}=-g_{\mu\nu},
\end{equation}
where $D_\mu$ and $N_\mu$ are the number of hole- and particle-states
 in species $\mu$, respectively.
The cases $g_{\mu\nu}=g\delta_{\mu\nu}$ with $g=0$ and $g=1$
correspond, respectively, to free bosons and fermions.
A remarkable point is that the statistical interaction
$g$ is uniquely determined by  the two-body 
phase shift in the Bethe equations. 
For instance, in  the  one-component system such 
as the SU(2) model, the statistical interaction $g$ is 
given by the two-body phase shift
 via the relation 
${1 \over 2}(g-1)\sum_{j(\ne i)}{\rm sgn}(k_i-k_j)$.
Therefore, we can deduce from the Bethe 
equations (\ref{BetAnsEqu}) for $\nu=2$  
 that the statistical interaction for 
the SU(2) Haldane-Shastry model is
$g=2$. Since the effect of the 
two-body phase shift is represented schematically by the motif,
one can  see that the above analysis based on the motif 
is directly  related to exclusion statistics.
Namely, the period for the spectral flow 
is determined by  the statistical interaction 
$g$. \footnote{If we include the Ising-anisotropy $\Delta=g(g-1)/2$
for the SU(2) model,
the period may become $g$, as was indeed shown for $g=4$
in\cite{FukKaw1}.}

For the SU($\nu$) Haldane-Shastry model, it is known that
the statistical interaction is
given by the $(\nu-1)\times(\nu-1)$ Cartan matrix 
of SU($\nu$) algebra\cite{Kaw},
\begin{equation}
\tilde g=\left(\begin{array}{rrrrr}2&-1&0&&\\-1&2&-1&0&\\
&&\ddots&&\\&0&-1&2&-1\\&&0&-1&2\end{array}\right), 
\label{CarMat}
\end{equation}
which directly reflects the structure of 
the nested Bethe equations (\ref{BetAnsEqu}).
We can easily generalize the above analysis to  
the SU($\nu$) model. For example, the period of the 
flow for $\phi_1=\phi_2=\cdots=\phi_{\nu-1}=\phi$
is $\nu$, which is related with
\begin{equation}
1-( \tilde g^{-1})_{11} = \frac{1}{\nu}.
\end{equation}
In this way, the period of the 
spectral flow in the SU($\nu$) model 
is determined by the matrix of the 
statistical interaction.
Note that 
the matrix $\tilde g$ corresponds to the spin sector of 
the topological-order 
matrix characterizing internal structure of the 
quantum Hall system \cite{topology}.

\section{Summary and Discussions}

Summarizing, we have proposed the SU($\nu$) Haldane-Shastry model
compatible with twisted boundary conditions, and have solved it
exactly. The spectrum thus obtained can be 
correctly reproduced  by the Bethe ansatz solution. We have then 
discussed the spectral flow of
the model in terms of the motif in the Bethe ansatz, 
and found that the period of the ground state is $\nu$,
reflecting fractional exclusion statistics.


Finally, we wish to make some comments on  an unusual 
linear  $\phi$-dependence of the spectral flow. 
We show here that this behavior is related to high 
symmetry inherent in  the present system.   
For this purpose let us first remember 
that the periodic Haldane-Shastry model has 
Yangian symmetry which is larger than SU($\nu$).
 Twisting the model breaks Yangian symmetry as well as
SU($\nu$). However, a nontrivial conserved quantity 
 associated with Yangian still
persists even for twisted systems of finite size\cite{FukKaw2}, 
and we expect that this conserved quantity 
controls  the linear dependence of the spectral flow.  
This feature can be clearly seen when we consider the conformal limit
of the model.
To be more specific,  let us
consider the SU(2) Haldane-Shastry model. 
For the periodic Haldane-Shastry model, 
we can take conformal limit  explicitly as follows\cite{HHTBP}:
\begin{equation}
H_{\rm PHS}=\sum_{n>0}n:J_{-n}^aJ_{n}^a:,
\end{equation}
where $J_n^a$ is the current operator of the $k=1$ SU(2) WZW model.
The level-1 Yangian generators are also given by\cite{HHTBP}
\begin{equation}
Q_1^a=\sum_{n>0}\epsilon^{abc}:J_{-n}^bJ_{n}^c: .
\end{equation}
For the twisted model, a similar calculation for the Hamiltonian
(\ref{FinHam}) leads to
\begin{equation}
H_{\rm THS}=H_{\rm PHS}+\phi Q_1^3 .
\label{nontri}
\end{equation}
From this equation, we can clearly see 
 that $ Q_1^3$ remains 
as a conserved quantity, reflecting high symmetry of the 
system, even when the boundary is 
twisted \cite{FukKaw2}. We can also see from
(\ref{nontri}) that the existence of this conserved quantity 
explains why we have encountered the  
linear $\phi$-dependence in the spectral flow.

We have been concerned with 
the  model of rational twist angles in this paper.
Within the present framework, we may apply
the continued fraction approximation 
to discuss  the case of irrational twist angles.
It remains still an open question to 
construct the exact solution directly for the 
irrational cases, which is now under consideration.
Also, the proof for integrability of the present model 
remains  as an interesting issue to be solved in the 
future work.

\acknowledgements
This work is partly supported by the Grant-in-Aid from the Ministry of
Education, Science and Culture, Japan.

\appendix
\section{}

In this appendix, we derive (\ref{ActTk}) in a bit detail.
We concentrate on the case where $Q_k\equiv N-M_k-M_0$ is an even
number and hence the current $J_k$ 
defined in eq.(\ref{Cur}) is an integer.
It is straightforward to apply a similar 
calculation to the other case ($Q_k=$ odd number). 

The key formula is
\begin{eqnarray}
S_{st}^\phi (J)&=&\frac{1}{4}
\sum_{n=1}^{N-1}J_\phi(n)z^{Jn}(1-z^n)^s(1-z^{-n})^t\nonumber\\
&=&\left\{\begin{array}{ll}
(-)^s& {\rm for}\quad s+t=2\\
(-)^s(J+\phi-\frac{N}{2})-\frac{1}{2}\quad& {\rm for}\quad s+t=1\\
2\varepsilon(J+\phi)+\frac{1}{3}(N^2-1)&{\rm for}\quad s=t=0 \\
0&{\rm for}\quad{\rm others},
\end{array}\right.
\label{SIde}
\end{eqnarray}
provided $t\le J+\phi\le N-s$,
which was first introduced in \cite{FukKaw1}, by extending the
original one in \cite{Hal}.
Now let us expand eq.(\ref{ActTk}) in power series of $1-z^n$ and
$1-z^{-n}$. It consists of finite polynomials of the form 
$(1-z^n)^s(1-z^{-n})^t$ with $0\le s+t\le Q_k/2+M_k-1$.
Therefore, provided that 
$Q_k/2+M_k-1\le J_k+\phi_k\le N-(Q_k/2+M_k-1)$,
terms with $3\le s+t$ should vanish in eq.(\ref{TBF}).
We can also find that  such terms in the second order as
$(1-z^n)^2-(1-z^{-n})^2$ or $\{(1-z^n)+(1-z^{-n})\}^2$ vanish,
so that it is sufficient to consider 
in the expansion
\begin{eqnarray}
\sum_\alpha\prod_{\beta(\ne\alpha)=1}^{M_k}B_{\alpha\beta}^{(n)}&&
\prod_{j=1}^{N-M_k-M_0}F_{\alpha j}^{(n)}\nonumber\\
&=&M_k-\sum_{\alpha\ne\beta}g_{\alpha\beta}^{(n)}+
\frac{1}{2}{\sum_{\alpha,\beta,\gamma}}'
g_{\alpha\beta}^{(n)}g_{\alpha\gamma}^{(n)}\nonumber\\
&-&\frac{Q_k}{8}(M_k-\sum_{\alpha\ne\beta}g_{\alpha\beta}^{(n)})
\{(1-z^n)+(1-z^{-n})\}\nonumber\\
&+&\frac{i}{4}\sum_\alpha(1-\sum_{\beta(\ne\alpha)}g_{\alpha\beta}^{(n)})
\{(1-z^n)-(1-z^{-n})\}\sum_j\cot\Theta_{\alpha j}\nonumber\\
&+&\frac{1}{4}\sum_\alpha(1-\sum_{\beta(\ne\alpha)}g_{\alpha\beta}^{(n)})
\{(1-z^n)+(1-z^{-n})\}\sum_{j\ne j'}\cot\Theta_{\alpha
j}\cot\Theta_{\alpha j'}.
\end{eqnarray}
Applying the formula (\ref{SIde}), we then have
\begin{eqnarray}
\frac{T_k(\phi_k)\psi}{\psi}
&=&2M_k\widetilde\varepsilon(J_k+\phi_k)+\frac{1}{3}M_k(N^2-1)+
\frac{2}{3}M_k(M_k^2-1)+\frac{1}{2}Q_kM_k\nonumber\\
&-&\sum_{\alpha\ne\beta}J_0(x_\alpha-x_\beta)\nonumber\\
&-&2i(J_k+\phi_k-N/2)\sum_{\alpha,j}\cot\Theta_{\alpha j}\nonumber\\
&-&2\sum_{\alpha\ne\beta,j}
\cot\Theta_{\alpha\beta}\cot\Theta_{\alpha j}
-\frac{1}{2}\sum_{i\ne j,\alpha}
\cot\Theta_{i\alpha}\cot\Theta_{j\alpha}.
\end{eqnarray}
So far we have used the simplified notations defined in
eq.(\ref{SimNot}). 
The remaining task is to convert the expression into the original
notation.  For example, in the third line, 
$\sum_{\alpha,j}\cot\Theta_{\alpha_kj}
=\sum_{\alpha_k}\sum_{l=1}^{\nu-1}\sum_{\alpha_l}
\cot\Theta_{\alpha_k\alpha_l}$.
Consequently we end up with the results shown in the text.



{\it Note added in proofs}

After submitted the paper, we have learned that Liu and Wang
(cond-mat/9608026) solved the supersymmetric $t$-$J$ model with 
$1/r^2$ interaction for irrational twist angles.
Their method can be directly applied to the present model.
 
\begin{figure}[h]
\epsfxsize=10cm
\centerline{\epsfbox{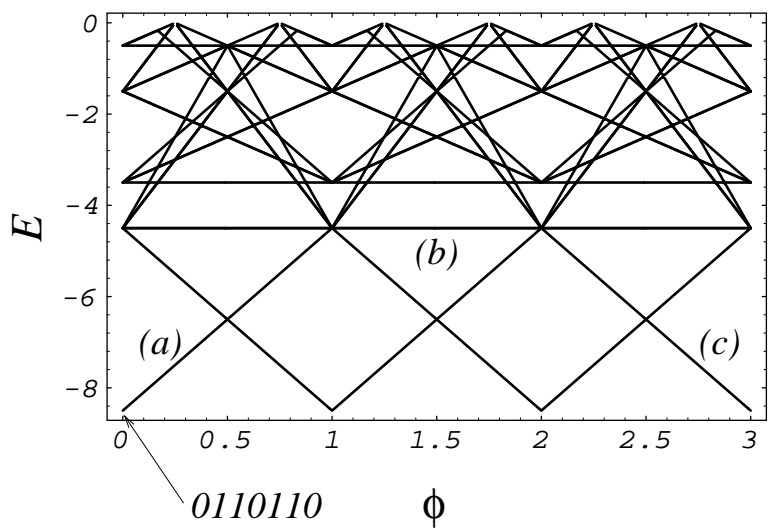}} 
\vspace{0.5cm}
\centerline{
\caption{Exact spectral flow of the $N=6$ SU(3) model with
$M_0=M_1=M_2=2$ and $\phi=\phi_1=\phi_2$.
Lower 20 levels are plotted.
The flow of the ground state (0110110) is described by 
$(a)\rightarrow (b) \rightarrow (c)$.
}
\label{fig:fig1}
}
\end{figure}


\begin{references}
\bibitem[\dagger]{Ema} email address: fukui@yukawa.kyoto-u.ac.jp
\bibitem[\ddagger]{Padd} Present address: Institute of Advanced Energy, Kyoto
University, Uji, Kyoto 611, Japan
\bibitem{Hal} F. D. M. Haldane, Phys. Rev. Lett. {\bf 60}, 635 (1988).
\bibitem{Sha} B. S. Shastry, Phys, Rev. Lett. {\bf 60}, 639 (1988).
\bibitem{ISMD1} F. Calogero, J. Math. Phys. {\bf 10}, 2197 (1969).
\bibitem{ISMD2} 
B. Sutherland, J. Math. Phys. {\bf 12}, 246, 256 (1971);
Phys. Rev. {\bf A4}, 2019 (1971); {\bf A5}, 1372 (1971).
\bibitem{HalFES} F. D. M. Haldane, Phys. Rev. Lett. {\bf 67}, 937
(1991).
\bibitem{FukKaw2} T. Fukui and N. Kawakami, 
Phys. Rev. Lett. {\bf 76}, 4242 (1996).
\bibitem{ABB} F. C. Alcaraz, M. N. Barber and M. T. Batchelar, 
Ann. Phys. {\bf 182}, 280 (1988).
\bibitem{ShaSut} B. S. Shastry and B. Sutherland, Phys. Rev. Lett.
{\bf 65}, 243 (1990);
B. Sutherland and B. S. Shastry, Phys. Rev. Lett. 
{\bf 65}, 1833 (1990).
\bibitem{Sut} B. Sutherland, Phys. Rev. Lett. {\bf 74}, 816 (1995).
\bibitem{FukKaw1}T. Fukui and N. Kawakami,
J. Phys. Soc. Jpn {\bf 65}, 2824 (1996).
\bibitem{KusAok} K. Kusakabe and H. Aoki, 
J. Phys. Soc. Jpn {\bf 65}, 2772 (1996). 
\bibitem{HaHal}Z. N. C. Ha and F. D. M. Haldane, Phys. Rev. 
{\bf B46}, 9359 (1992).
\bibitem{Kaw} N. Kawakami, Phys. Rev. {\bf B46}, (1992) 1005.
\bibitem{HHTBP} F. D. M. Haldane, Z. N. C. Ha, J. C. Talstra, D. Bernard
 and V. Pasquier, Phys. Rev. Lett. {\bf 69}, 2021 (1992).
\bibitem{WLC} D. F. Wang, J. T. Liu and P. Coleman, 
  Phys. Rev. {\bf B46}, 6639 (1992).
\bibitem{FukKawTJ} T. Fukui and N. Kawakami,
Phys. Rev. {\bf B54}, 5346 (1996).
\bibitem{sunsuth} B. Sutherland, Phys. Rev. {\bf B12}, 3795 (1975).
\bibitem{topology} B. Blok and X. G. Wen, Phys. Rev. {\bf B42},
8133 (1990); N. Read, Phys. Rev. Lett. {\bf 65}, 1502 (1990).
\end{references}
\end{document}